\documentstyle[twocolumn,aps]{revtex}

\newcommand{\be}{\begin{equation}}
\newcommand{\ee}{\end{equation}}
\newcommand{\ben}{\begin{eqnarray}}
\newcommand{\een}{\end{eqnarray}}

\newcommand{\iii}{\'{\i}}

\newcommand{\nd}{\noindent}

\begin{document}
\title{Maximally entangled mixed states and conditional entropies}
\author{J. Batle$^1$,  M. Casas$^1$,
 A. Plastino$^{2,\,3}$, and A. R. Plastino$^{3,\,4,\,5}$}

\address {
$^1$Departament de F\iii sica, Universitat de les Illes Balears and IMEDEA-CSIC,
07122 Palma de Mallorca, Spain \\
  $^2$Argentina's National Research Council (CONICET) \\
$^3$Department of Physics, National University La Plata,
  C.C. 727, 1900 La Plata, Argentina\\
$^4$Faculty of Astronomy and Geophysics, National University La Plata,
  C.C. 727, 1900 La Plata \\
$^5$Department of Physics, University of Pretoria,
 0002 Pretoria, South Africa
}

\date{\today}

\maketitle
\begin{abstract}

\nd  The maximally entangled mixed states of Munro, James, White, and
Kwiat [Phys. Rev. A {\bf 64} (2001) 030302] are shown to exhibit
interesting features vis a vis conditional entropic measures. The
same happens with the Ishizaka and Hiroshima states [Phys. Rev. A
{\bf 62} 022310 (2000)], whose entanglement-degree can not be
increased by acting on them with logic gates. Special types of  
entangled states that do not violate classical entropic inequalities are seen to
exist in the space of two qubits. Special meaning can be assigned
to the Munro {\it et al.} special participation ratio of 1.8.

\noindent
 Pacs: 03.67.Mn; 89.70.+c

\end{abstract}

\maketitle


\section {Introduction}  Entanglement is one of the
most fundamental issues of quantum theory \cite{LPS98}. It is a
physical resource, like energy, associated with the peculiar
non-classical correlations that are possible between separated
quantum systems. Recourse to entanglement is required so as  to
implement quantum information processes \cite{W98,BEZ00} such as
quantum cryptographic key distribution \cite{E91}, quantum
teleportation \cite{BBCJPW93}, superdense coding \cite{BW93}, and
quantum computation \cite{BDMT98}. Indeed, production of
entanglement is a kind of elementary prerequisite for any quantum
computation.    A state of a composite quantum system is called
``entangled" if it can not be represented as a mixture of
factorizable pure states. Otherwise, the state is called
separable. The above definition is physically meaningful because
entangled states (unlike separable states) cannot be prepared
locally by acting on each subsystem individually \cite{We89,P93}.
A physically motivated measure of entanglement is
 provided by the entanglement of formation $E[\rho]$  \cite{BDSW96}, that
  quantifies the resources needed to create a given entangled state $\rho$.
  The entanglement of formation for two-qubits
 systems is given by Wootters' expression \cite{WO98},
$E[\rho] \, = \, h\left(1+\sqrt{1-C^2}/2\right)$, 
where $ h(x) \, = \, -x \log_2 x \, - \, (1-x)\log_2(1-x), $
\noindent and $C$ stands for the {\it concurrence}  of the
two-qubits state $\rho$. The concurrence is given by $ C \, = \,
max(0,\lambda_1-\lambda_2-\lambda_3-\lambda_4),$ \noindent
$\lambda_i, \,\,\, (i=1, \ldots 4)$ being the square roots, in
decreasing order, of the eigenvalues of the matrix $\rho \tilde
\rho$, with $ \tilde \rho \, = \, (\sigma_y \otimes \sigma_y)
\rho^{*} (\sigma_y \otimes \sigma_y)$. The above expression has to
be evaluated by recourse to the matrix elements of $\rho$ computed
with respect to the product basis. \nd Another 
meaningful quantity is the fully entangled fraction $F_{EF}$ 
\cite{j02}, that determines the range of possible 
concurrence values for a mixed state: $F_{EF} \le C \le (F_{EF}+1)/2$.  
 For an illustration of this last
statement, the reader is referred to  Fig. 2 of Ref. \cite{j02}, whose authors 
  investigate the fraction of two-qubits
mixed states that can be used in all quantum information processing 
applications using  $F_{FE}$.
 Still another important quantity is the participation ratio,

  \be \label{partrad}
  R(\rho) \, = \, [Tr(\rho^2)]^{-1},
  \ee

\noindent is particularly convenient for calculations and can be regarded as a 
measure of the degree of mixture of a given density matrix 
\cite{ZZF00,ZHS98,MJWK01}. It varies from unity for pure states to $N$ 
for totally mixed states (if $\hat \rho$ is represented by 
an $N \times N$ matrix). It may be interpreted as 
the effective number of pure states that enter the mixture. If the 
participation ratio of $\rho$ is high enough, then its partially 
transposed density matrix is positive, which for $N=4$ amounts to 
separability for $R \ge 3$ \cite{P93,ZHS98}. Notice also that $R$ is invariant 
under the action of unitary operators.

   There are several entropic (or information) measures that can
  be useful in order to investigate the violation of classical entropic
  inequalities by quantum entangled states. Among them,
the von Neumann measure
 is important because of its relationship with the thermodynamic
  entropy, and the participation
  ratio  is particularly convenient both for numerical
and analytical calculations \cite{ZZF00,ZHS98,MJWK01}.
  The $q$-entropies, which are functions of the quantity $
  \omega_q \, = \, Tr \left( \rho^q \right),$
    provide one with a whole family of entropic measures.
  In the limit $q\rightarrow 1 $ these measures incorporate von
  Neumann's as a particular instance. On the other hand, when $q=2$ they are
  simply related to the participation ratio (\ref{partrad}). Most of the
  applications of $q$-entropies to physics involve either the R\'enyi
  or the Tsallis' entropies \cite{casas,gellmann}, respectively,

   \be \label{renyi}
   S^R_q \, = \, \ln \left( \omega_q
   \right)/(1-q);\,\,\,\,
   S^T_q \, = \, \bigl(1-\omega_q \bigr)/(q-1).
  \ee
    \noindent In the  $q=2-$case, $ S^T_{q=2}$ is often called
  {\it the linear entropy} $\mathcal{S}_L$  \cite{MJWK01}. 
%
  Tsallis' and R\'enyi's measures are
  related through $S^T_q \, = \,F( S^R_q),$ where the function $F$ 
is given by $F(x)  =   \left\{ e^{(1-q)x} - 1 \right\}/(1-q).$
  As an immediate consequence, for all non vanishing values of $q$, 
Tsallis' measure $ S^T_q$ is a monotonic increasing function of R\'enyi's
  measure $ S^R_q $.
  Considerably attention has been recently paid to a conditional
  entropic measure based upon Tsallis' functional, and defined as

  \be \label{qurela}
  S^T_q(A|B) \, = \,
  \{S^T_q(AB)-S^T_q(B)\}/\{1+(1-q)S^T_q(B)\}.
  \ee
  \noindent
  Here $\rho_{AB}$ designs an arbitrary quantum state of the
  composite system $A\oplus B$,
  not necessarily factorizable nor separable,
  and $\rho_B = Tr_A (\rho_{AB})$. The conditional $q$-entropy
  $S^T_q(B|A)$ is defined in a similar way as (\ref{qurela}),
  replacing $\rho_B $ by $\rho_A = Tr_B (\rho_{AB})$.
     The conditional $q$-entropy
  (\ref{qurela}) has been recently studied in connection
  with the separability of density matrices describing composite
  quantum systems \cite{TLB01,TLP01}. For separable states
   (see for instance \cite{usEPJ03})

  \be \label{qsepar}
  S^T_q(A|B) \ge  0; \,\,\,\,  S^T_q(B|A) \ge  0.
  \ee
  \noindent
  On the contrary, there are entangled states that have negative
  conditional $q$-entropies. That is, for some entangled states one (or both)
  of the inequalities (\ref{qsepar}) are not verified.
Now, since Tsallis' entropy is a monotonous
  increasing function of R\'enyi's,
  it is plain that (\ref{qurela}) has always the same sign as
$S^R_q(A|B) = S^R_q(\rho_{AB})-S^R_q(\rho_{B}).$
  The positivity of either
 the Tsallis' conditional entropy or the R\'enyi
 conditional entropy  are known as the ``classical $q$-entropic
 inequalities" \cite{usEPJ03}.

In practice, one will
more often have to deal with mixed states than with pure ones.
From the point of view of entanglement-exploitation, one should
then be interested in maximally entangled mixed states (MEMS)
 $\rho_{MEMS}$, which are the basic constituents of all quantum communication 
protocols. The MEMS states have been studied, for example,
in Refs. \cite{MJWK01,IH00,W03} for the two-qubits instance of two (one
qubit-)subsystems $A$ and $B$. For MEMS, the relations between i) von
  Neumann's and linear entropies, on the one hand, and ii) concurrence 
and von Neumann entropy, on the other one, 
have been exhaustively investigated in \cite{W03}.  
MEMS states have been recently been experimentally 
encountered \cite{memsexp,memsexp2}. We will focus attention on these 
kind of states here. 
MEMS for a given $R-$value have the following appearance in
the computational basis
($|00\rangle,|01\rangle,|10\rangle,|11\rangle$) \cite{MJWK01}.

\be \rho_{MEMS} = \left( \begin{array}{cccc}
g(x) & 0 & 0 & x/2\\
0 & 1 - 2g(x) & 0 & 0\\
0 & 0 & 0 & 0\\
x/2 & 0 & 0 & g(x) \end{array} \right), \label{MEMS} \ee 

\noindent with $g(x)=1/3$ for $0\le x \le 2/3$, and  $g(x)=x/2$ for $2/3 \le x
\le 1$. The change of $g(x)-$regime ensues for $R=1.8$. {\it We
will reveal below some physical consequences of this
regime-change} . Of great importance are also mixed states whose
entanglement-degree cannot be increased by the action of logic
gates \cite{IH00} that, again in the same basis, are given by

\begin{equation} \label{IH}
\rho_{IH} = \left( \begin{array}{cccc}
p_2 & 0 & 0 & 0\\
0 & \frac{p_3+p_1}{2} & \frac{p_3-p_1}{2} & 0\\
0 & \frac{p_3-p_1}{2} & \frac{p_3+p_1}{2} & 0\\
0 & 0 & 0 & p_4 \end{array} \right),
\end{equation}

\noindent whose eigenvalues are the $p_i;\,\,(i=1,\ldots,4)$ and   $p_1 \ge p_2 \ge p_3
 \ge p_4$. We call these states the Ishizaka and Hiroshima (IH) ones and their
concurrence reads $C_{IH}\,=\,p_1\,-\,p_3\,-\,2\,\sqrt{p_2\,p_4}$, 
a relation valid for ranks $\le 3$ that has numerical support also
if the rank is four \cite{IH00}. Of course, all MEMS belong to the
IH-class. Our goal is to uncover interesting correlations between
entanglement and mixedness that emerge when we study these states
from the view point of conditional entropies.

\section {Entropic inequalities and MEMS} 

We begin here with the presentation
of our results. A few of them are of an analytical nature. For
instance, in the case of all states of the forms (\ref{IH}) and/or
(\ref{MEMS}), the partial traces $\rho_{A/B}$ over one of the
subsystems $A$ or $B$ are equal, i.e., for the reduced density
matrices we have $\rho_A=\rho_B$, which entails
$S_{q}(A|B)=S_{q}(B|A)$ for both the R\'enyi and the Tsallis entropy. 
Notice that this is a particular feature of
these states.

\nd As for the form (\ref{IH}), we establish a lower bound to its
states' concurrence for a considerable $R-$range 
(see Fig. 3),     
namely, \be
\label{lowerb} C_{IH;Min}=[\sqrt{3R(4-R)}-R]/(2R).\ee

\nd In the case of MEMS and in the vicinity of $R=1$ we can
analytically relate entropic changes with concurrence-changes, in
the fashion (remember that for MEMS $C \equiv C_{Max}$)
\begin{equation} \label{relaS}
\Delta S^R_{q}(A|B)\,=\,-[2q/\{\ln(2)(q-1)\}]\,\Delta C.
\end{equation}
The case $q \rightarrow \infty$ is the strongest $q$-entropic
criterion \cite{usEPJ03}.
 Eq. (\ref{relaS}) expresses the fact that, for MEMS, small deviations
 from pure states (for which the $q$-entropic criteria are necessary and
sufficient separability conditions) do not change the criteria's
validity, that becomes then {\it extended} to a class of mixed
states.  
\subsubsection{Numerical results}
We will randomly generate states in the space ${\cal S}^{(N)}$ of mixed 
states $\rho$ ($N=4$ in our case).  This 
can be regarded as a product space,  
${\cal S}^{(N)}={\cal P}\times \Delta$, 
where ${\cal P}$ stands for the set of orthonormal projectors 
($\sum_{i=1}^{N} \hat P_i = I$) and $\Delta$ is the set of all real 
$N-$uples 
($\{\lambda_i\}, \, 0 \le \lambda_i \le 1, \, \sum_{i=1}^N \lambda_i=1$). All 
states $\rho$ are generated according to the ZHSL measure 
$\nu\times {\cal L}_{N-1}$. Here, $\nu$ is the measure induced 
on ${\cal P}$ by the Haar measure on the 
group of unitary matrices $U(N)$ and ${\cal L}_{N-1}$ is the 
Leguesbe measure on the simplex of eigenvalues $\Delta$ \cite{Z99,BCPP02a}. 

As stated above, we deal in this
paper with two kinds of maximally entangles states (MEMS and
Ishizaka and Hiroshima ones). We call the class that comprises
both kinds the ME-one. Fig. 1 depicts the overall situation. In
the upper part we plot the ME-states' concurrence $C_{IH}$ vs.
the participation ratio. $R$ ranges in the interval $1 < R < 1.8$
(the latter figure corresponds to the above mentioned transition
point for MEMS). (A): the upper line gives MEMS-states and the
inferior one  the lower bound (\ref{lowerb}). (B): the lower part
of the Figure gives the conditional entropy of the ME states
$S^R_{q}(A|B)$ for $q \rightarrow \infty$ 
(the solid curve corresponds to the MEMS case).
It is always negative, so that here 
the entropic inequalities provide the correct aswer in order to 
detect entanglement.

Fig. 2 is a plot of the concurrence $C_{IH}$ vs.
$\lambda_{max}$, the maximum eigenvalue of our ME bipartite states
$\rho$. The dashed line corresponds to MEMS. The graph confirms
the statement made in  \cite{MJWK01} that the latter are not
maximally entangled states if mixedness is measured according to a
criterion that is not the $R-$one. Three separate regions (I, II,
III) can be seen to emerge. The maximum and minimum (continuous)
contour lines are of an analytical character:
\begin{itemize} \item {\sf First zone:}
a) $C_{IH}^{max}= \lambda_{max}$ for $\lambda_{max} \in [1/2, 1]$
\item b) $C_{IH}^{min}= 2\lambda_{max}-1 $ for Bell diagonal states.

\item {\sf Second:} a) $C_{IH}^{max}= 3\lambda_{max}-1$ for $\lambda_{max} \in [1/3, 1/2]$
\item b) $C_{IH}^{min}= 0$
\item  {\sf Third:} All states are separable $C_{IH}= 0$.
\end{itemize}
\nd Our three zones  (I, II, III) can be characterized according
to strict geometrical criteria, as extensively discussed in
\cite{BCPP02b}.
In point of fact, the paper by  Wei {\it et al.} \cite{W03} exhaustively 
studies MEMS for different measures of entanglement and mixedness. The extension 
made here to $\lambda_{max}$ as a proper degree of mixture confirms in 
Fig. 2 the discussion given in \cite{W03} that asserts that MEMS are sensitive 
to the form of mixture employed. 

Fig. 3 is a $C_{IH}$ vs. $R$ plot like that of 
Fig. 1, but for an extended $R-$range ($1<R<3$). The pertinent IH 
bipartite states fill a ``band" with dots (a sample of $10^4$ states). 
In Fig. 3 we focus
attention on a special type of bipartite states: those that, being
entangled, do fulfill the inequalities (\ref{qsepar}). {\it For these states}, 
let us call them entangled states with
\lq\lq classical'' {\it conditional} entropic behavior (ESCRE), 
the quasi-triangular
solid line depicts, for each $R$, the maximum degree of
entanglement attainable. 
For each value of $R$ (crosses), we generate 
$10^8$ states according to the aforementioned ZHSL measure, keeping only 
the ESCRE ones with maximal $C$.  Interestingly enough, 
{\sf the maximum
degree of entanglement for ESCRE obtains at} $R=1.8$, which
signals the change of regime for MEMS (Cf. (\ref{MEMS}) and
commentaries immediately below that equation). {\it This fact
gives an entropic meaning to that particular $R-$value}.  We can state then
that
i) whenever the entropic criterium turns out to
constitute a necessary and sufficient condition for separability (at
$R=1$ and $R=3$), the ESCRE-degree of entanglement is null, and
ii) the ESCRE-degree of entanglement is maximal at the Munro {\it et
al.} change-of-regime $R-$value of 1.8. 

\section {\bf Conclusions} 

For entangled states with classical conditional entropic behavior
(ESCRE),  the
maximum degree of entanglement attainable obtains at $R=1.8$.
Even though the entropic criteria are not universally valid for all
two-qubits states (yielding only a necessary condition for
separability), they have been shown here  to preserve their full
applicability for an important family of states, namely, those with
cannot increase their entanglement under the action of logic gates
for participation rations in the interval ($R\in[1,1.8]$). This in
turn, gives an entropic meaning to this special $R-$value
encountered by Munro {\it et al.} \cite{MJWK01}.
We find explicit ``boundaries" to $C_{IH}$ when we express the
degree of mixture using the maximum eigenvalue $\lambda_{max}$ of
$\rho^{IH}$. It would seem  that the characterization of the
entanglement for these states, using the $\lambda_{max}$ criterion,
provides the best insight into the entanglement features of these
states.
Beyond a certain value of the concurrence,
{\it all states}, not necessarily the ones considered before, can be
correctly described by the entropic inequalities as far as this
criterion is concerned. One may argue that if the quantum
correlations are strong enough (greater than $C^{max}_{R=1.8}$ or
$C^{max}_{\lambda_{max}=\frac{2}{3}}$), there is still room for
entropic-based separability criteria to hold.

\vskip 2mm {\bf Acknowledgements} \vskip 2mm
This work was partially supported by the MEC grant BFM2002-03241 (Spain)
and FEDER (EU), by the Government of Balearic Islands and by CONICET
(Argentine Agency).

\vskip 0.5cm
{\bf FIGURE CAPTIONS}

\noindent Fig. 1- Plot of the concurrence $C_{IH}$ for  two kinds
of maximally entangled states: Ishizaka and Hiroshima ones (dots) and
MEMS vs. $R$ (upper solid curve), for a sample-set. Their corresponding
$S^R_{\infty}(A|B)$-values are also shown. Contour lines can be found
analytically. See text for details. 

\vskip 0.5cm

\noindent Fig. 2- Plot of the concurrence $C_{IH}$ for the class of maximally
entangled states vs. their maximum eigenvalue $\lambda_{max}$ for a sample set
of states. The dashed line corresponds to $\rho_{MEMS}$-states. Notice the
fact that these states are not maximally entangled if the mixedness is not given
by $R$. Maximum and minimum contour-lines for $C_{IH}$ are found 
in analytical fashion. See text for details.

\vskip 0.5cm

\noindent Fig. 3- Same as in Fig. 1, but for  an extended $R$-range.
The lower curve (with crosses on it) represents, for each $R$-value, 
the maximum concurrence
 for those states which obey classical entropic
inequalities. The curve exhibits a maximum at
$R=1.8$ and it is vanishes at $R=1$ and $R= 3$, where the entropic criterion is
necessary and sufficient. That this curve does not exactly match the
MEMS \lq\lq quasi-diagonal''curve above it, for the range $[1.8,3)$,  
is due to the
relative scarcity of the pertinent states (generated randomly
according to the ZHSL measure). See text for details.

\enddocument
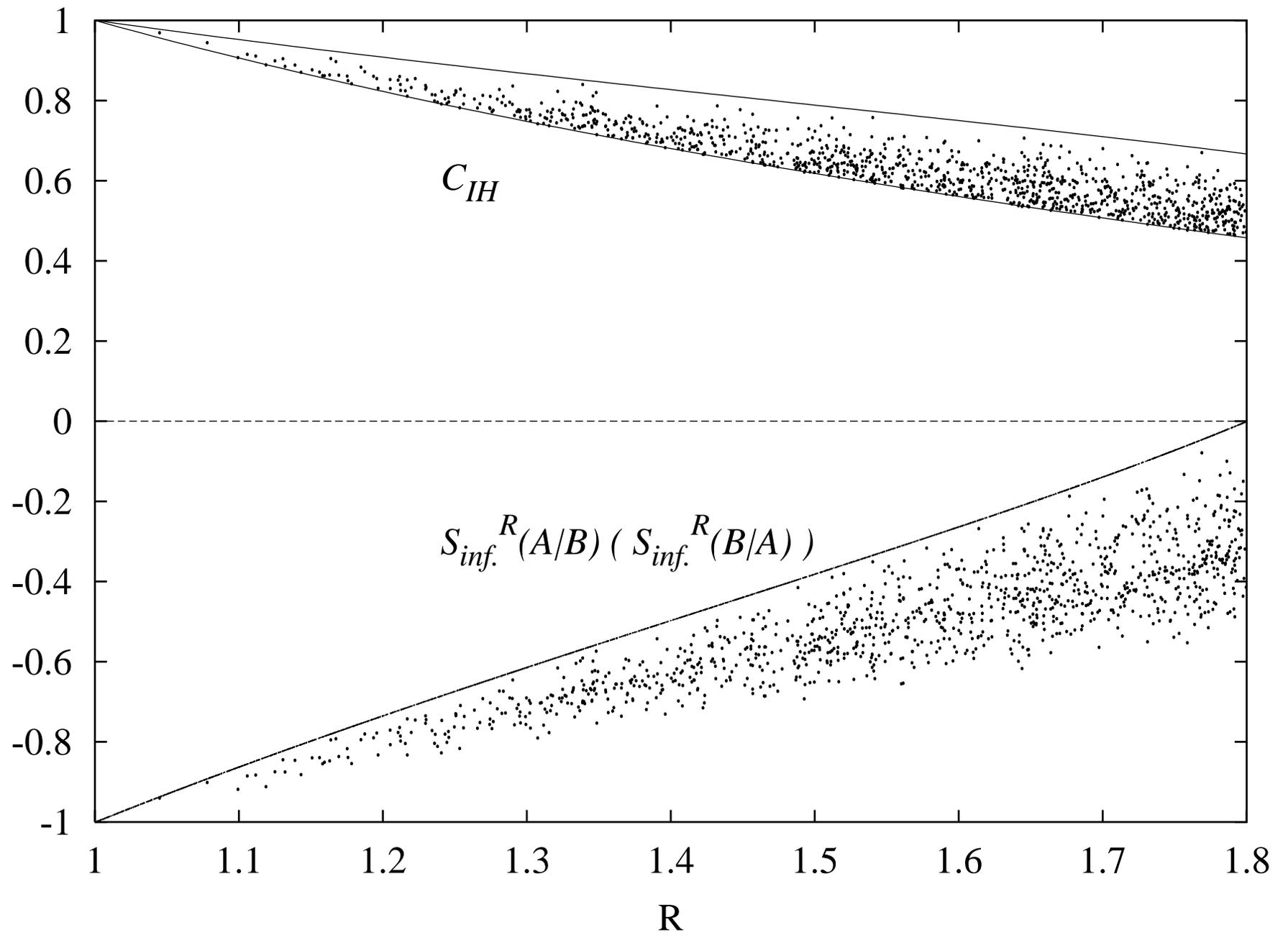
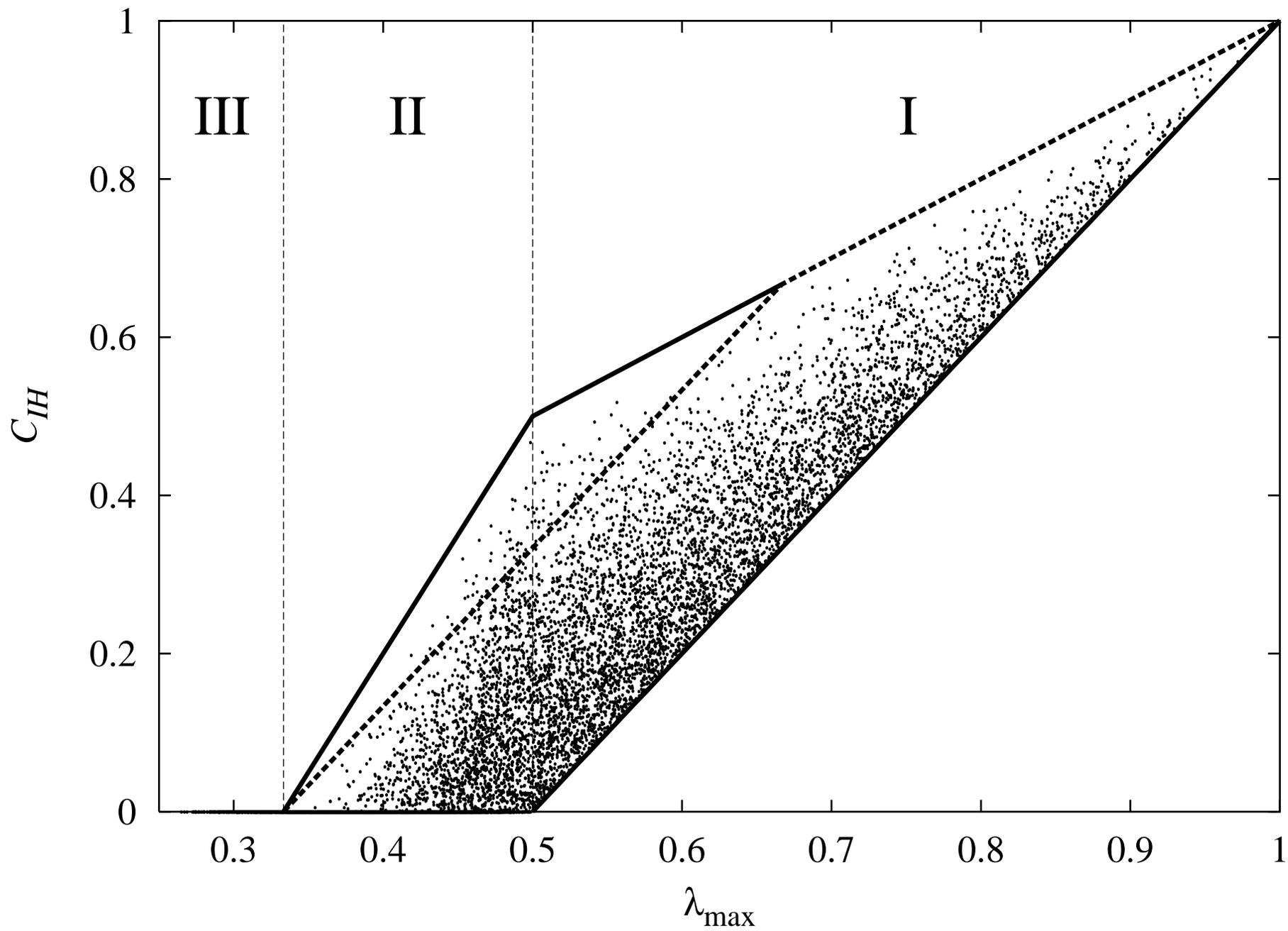
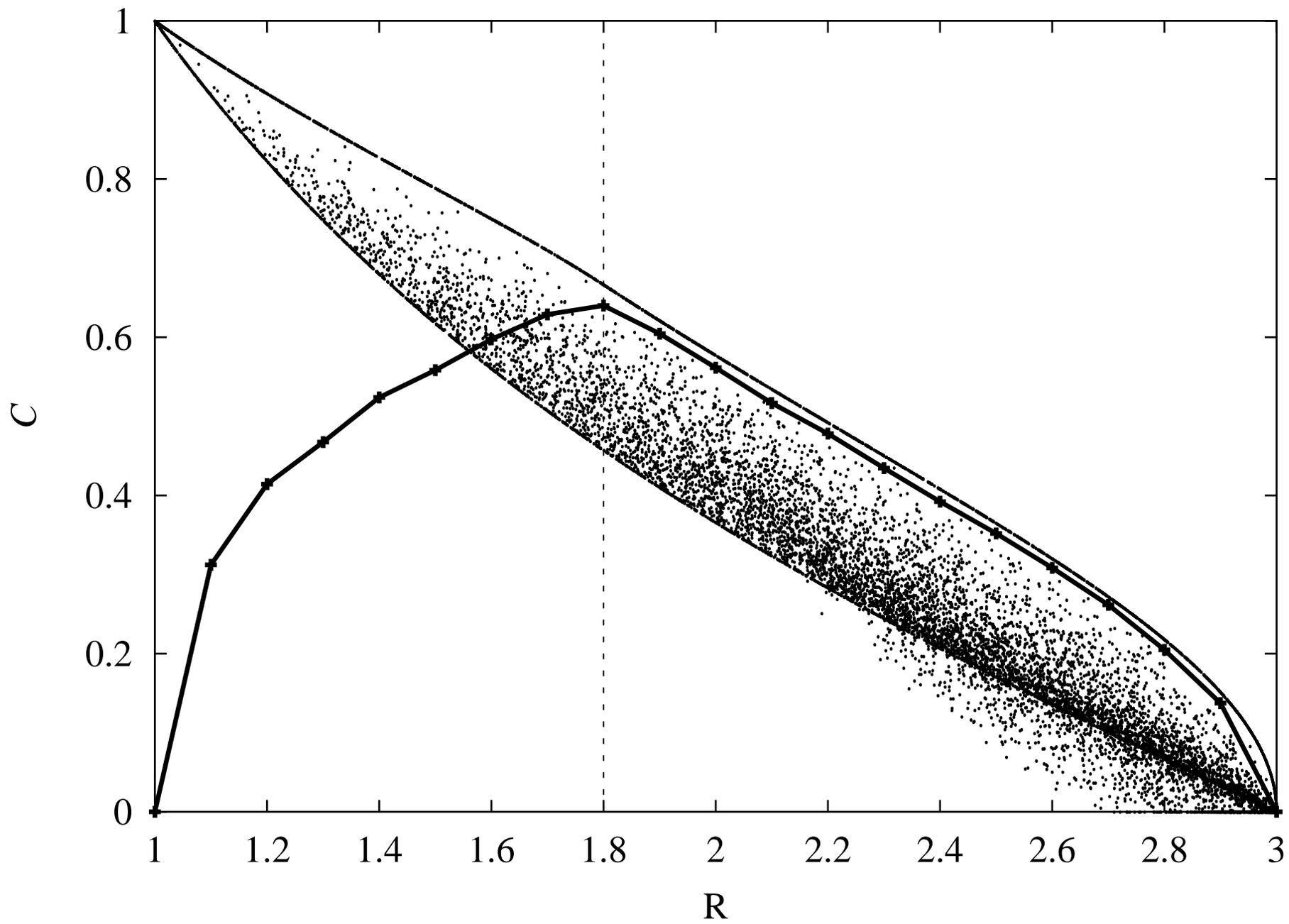
